\documentclass[twocolumn]{aastex62}

\newcommand{\mincir}{\ \raise -2.truept\hbox{\rlap{\hbox{$\sim$}}\raise5.truept   \hbox{$<$}\ }}
\newcommand{\magcir}{\ \raise -2.truept\hbox{\rlap{\hbox{$\sim$}}\raise5.truept
    \hbox{$>$}\ }}

\graphicspath{{./}{figures/}}

\received{..., 2019}
\revised{... 2019}
\accepted{...}
\submitjournal{ApJ}

\shorttitle{Finding the brightest cosmic beacons in the Southern Hemisphere}
\shortauthors{Calderone et al.}

\newcommand{\nmain}[0]{1,014,875}
\newcommand{\nveron}[0]{4,447}
\newcommand{\nsdss}[0]{1,365}
\newcommand{\ntwodfgrs}[0]{4,032}
\newcommand{\nconfirmed}[0]{4,666}
\newcommand{\nstars}[0]{844,026}
\newcommand{\ngal}[0]{3,665}
\newcommand{\nunk}[0]{162,518}
\newcommand{\nknown}[0]{852,357}
\newcommand{\nfinal}[0]{1476}
\newcommand{\nOBSfin}[0]{70}
\newcommand{\nADD}[0]{67}
\newcommand{\nFINconf}[0]{56}
\newcommand{\CCA}{\texttt{CCA}}

\newcommand{\OVI} {\mbox{O\,{\sc vi}}}
\newcommand{\Lya} {\mbox{Ly-$\alpha$}}
\newcommand{\CIV} {\mbox{C\,{\sc iv}}}
\newcommand{\HeII} {\mbox{He\,{\sc ii}}}

\begin{document}

\title{Finding the brightest cosmic beacons in the Southern Hemisphere}

\correspondingauthor{Giorgio Calderone}
\email{giorgio.calderone@inaf.it}

\author[0000-0002-7738-5389]{Giorgio Calderone}
\affil{INAF--Osservatorio Astronomico di Trieste\\
Via G.B. Tiepolo, 11, I-34143 Trieste, Italy \\}

\author[0000-0003-4432-5037]{Konstantina Boutsia}
\affil{Las Campanas Observatory, Carnegie Observatories, \\ 
Colina El Pino, Casilla 601, La Serena, Chile\\}

\author[0000-0002-2115-5234]{Stefano Cristiani}
\affil{INAF--Osservatorio Astronomico di Trieste\\
Via G.B. Tiepolo, 11, I-34143 Trieste, Italy \\}
\affiliation{INFN-National Institute for Nuclear Physics \\ 
via Valerio 2, I-34127 Trieste \\}
\affil{IFPU--Institute for Fundamental Physics of the Universe\\ via Beirut 2, I-34151 Trieste, Italy}

\author[0000-0002-5688-0663]{Andrea Grazian}
\affil{INAF--Osservatorio Astronomico di Padova \\
Vicolo dell'Osservatorio 5, I-35122, Padova, Italy\\}

\author[0000-0001-5758-1000]{Ricardo Amorin}
\affil{Instituto de Investigaci\'on Multidisciplinar en Ciencia y Tecnolog\'ia, Universidad de La Serena\\
Raul Bitr\'an 1305, La Serena, Chile}
\affil{Departamento de F\'isica y Astronom\'ia, Universidad de La Serena\\
Av. Juan Cisternas 1200 Norte, La Serena, Chile}

\author[0000-0003-3693-3091]{Valentina D'Odorico}
\affil{INAF--Osservatorio Astronomico di Trieste\\
Via G.B. Tiepolo, 11, I-34143 Trieste, Italy \\}
\affiliation{Scuola Normale Superiore\\
P.zza dei Cavalieri, I-56126 Pisa, Italy\\}

\author[0000-0002-6830-9093]{Guido Cupani}
\affil{INAF--Osservatorio Astronomico di Trieste\\
Via G.B. Tiepolo, 11, I-34143 Trieste, Italy \\}

\author[0000-0003-4744-0188]{Fabio Fontanot}
\affil{INAF--Osservatorio Astronomico di Trieste\\
Via G.B. Tiepolo, 11, I-34143 Trieste, Italy \\}
\affil{IFPU--Institute for Fundamental Physics of the Universe\\ via Beirut 2, I-34151 Trieste, Italy}

\author{Mara Salvato}
\affil{Max-Planck-Institut f{\"u}r extraterrestrische Physik\\
Giessenbachstrasse 1, Garching D-85748, Germany\\}

\begin{abstract}
  The study of absorptions along the lines of sight to bright high-$z$ QSOs is an invaluable cosmological tool that provides a wealth of information on the inter-/circum-galactic medium, Dark Matter, primordial elements, reionization, fundamental constants, and General Relativity. Unfortunately, the number of bright ($i \lesssim$~18) QSOs at $z \magcir 2$ in the Southern hemisphere is much lower than in the North, due to the lack of wide multi-wavelength surveys at declination $\delta <$~0$^\circ$, hampering the effectiveness of observations from southern observatories. In this work we present a new method based on Canonical Correlation Analysis to identify such objects, taking advantage of a number of available databases: Skymapper, Gaia DR2, WISE, 2MASS. Our QSO candidate  sample lists \nfinal{} sources with $i < 18$
 over 12,400 square degrees in the southern hemisphere.  With a preliminary campaign we  observed spectroscopically \nOBSfin{} of them, confirming \nFINconf{} new bright QSOs at $z > 2.5$, corresponding to a success rate of our method of $\sim$~80\%.
 Furthermore, we estimate a completeness of $\sim$~90\% of our sample at completion of our observation campaign.  The new QSOs confirmed by this first and the forthcoming  campaigns will be the targets of subsequent studies using higher resolution spectrographs, like ESPRESSO, UVES, and (in the long term) ELT/HIRES.
\end{abstract}


\section{Introduction} \label{sec:intro}
The study of absorption lines in the spectra of high redshift Quasi-Stellar Objects (QSO) is a fundamental tool for Cosmology \citep{meiksin09, mcquinn16}.  Along the lines of sight to these powerful light beacons, every parcel of the intervening gas selectively absorbs wavelengths of light, providing information about the spatial distributions, motions, temperature, chemical enrichment, and ionization histories of gaseous structures from redshift seven and beyond until the present.

In particular, thanks to QSO absorption lines, it is possible to address issues like: What were the physical conditions of the primordial Universe? What fraction of the matter was in a diffuse medium and what fraction and how early condensed in clouds?  Where are most of the baryons at the various redshifts?  When and how did the formation of galaxies and large scale structure start?  How early and in what amount have metals been produced?  When and how (after the Dark Ages following recombination) did the Universe get re--ionized?  What was the typical radiation field, how homogeneous, and what was producing it?  Which constraints on cosmological parameters and types of dark matter (e.g. neutrinos) are derived from the large scale structure traced by the inter-galactic medium (IGM)?  Does the standard Big Bang nucleosynthesis model makes the correct predictions about the primordial element abundances and the temperature evolution of the CMB?  Do fundamental constants of Physics (e.g. the fine structure constant, $\alpha$, or the proton-to-electron mass ratio, $\mu$) vary with cosmic time?  Does General Relativity correctly describe the expansion of our Universe? 
In order to efficiently pursue these and other similar lines of investigation, it is essential to have the brightest possible light beacons in the background. 

Historically, observations in the Southern hemisphere have been hampered by the lack of luminous targets with respect to the North, due to a lesser investment of telescope time to search for bright QSOs in the South. As an example, the {\it Quasar Deep Spectrum} observations carried out with the UVES spectrograph \citep{dodorico16} have targeted the QSO HE0940-1050 (z$_{\rm em} = 3.09$, V=16.9) and required 64.4 hours to reach a Signal-to-Noise-Ratio (SNR), per resolution element ($R = 45,000$), of 120-500 and 320-500 in the \OVI{}/\Lya{} region and in the \CIV{} region, respectively.  HE0940-1050 is still the best target at this redshift in the South, but it is not comparable to the (lensed) beacons B1422+231 (z$_{\rm em} = 3.62$, V=15.8) or APM 08279+5255 (z$_{\rm em} = 3.91$, V=15.2), which have been available for observers in the Northern Hemisphere.

It is particularly urgent now to fill this gap in view of the upcoming new instrumentation in the Southern hemisphere, like ESPRESSO at VLT and the planning of new experiments (e.g. the Sandage test with the high-resolution spectrograph HIRES at the E-ELT \citep{cristiani07, liske08}, or the test of the stability of the fine-structure constant and other fundamental couplings \citep{leite16}.
Moreover, finding bright radio-loud QSOs at high-$z$ is particularly important to study the 21cm forest in absorption with future breakthrough facilities, like the Square Kilometer Array (SKA) in the Southern hemisphere, as proposed by \citet{Carilli02}. In addition, UV/optically bright QSOs at $z>3$ with lines of sight free from Lyman Limit Systems (LLS) up to the \HeII{} forest are particularly rare but extremely valuable to study the \HeII{} Reionization \citep{syphers14, worseck16, worseck19}.

By Comparing QSO surface densities, it is statistically evident that relatively high-$z$ objects of bright apparent magnitudes must be also present in the Southern hemisphere: of the 22 known QSOs with $z>2.8$ and $V<17$, only 5 are at $\delta < 0^\circ$, and all the 3 with $V<16$ are in the North.
Such unbalance exists because historically many surveys (e.g. the SDSS) have focused their efforts mainly in the Northern hemisphere. The present time however is ripe for a dramatic change of scenario thanks to new surveys available through the whole sky, or insisting mainly in the Southern hemisphere, such as Gaia DR2, Skymapper, 2MASS, and WISE \citep{RefGaiaMain, RefGaiaDR2, RefSkymapper, Ref2MASS, RefWise}.

In this paper we describe the first results of a program aiming at filling this gap in the Southern hemisphere, finding the brightest QSOs at $z>2.5$ that will be observed at high resolution with the present and future breakthrough facilities.

\begin{deluxetable*}{|l|c|c|c|c|c|}
  \tablecaption{Cumulative surface density of QSOs at different redshifts and $i$-band magnitude limits expected from the luminosity function of \citet{kulkarni18}.  The minimum and maximum range of surface density in a given redshift interval for QSOs brighter than a given $i$-band magnitude limit is provided.  The expected bright QSO number counts are based on the best fit of individual luminosity functions by \citet{kulkarni18} (their Table 2), as well as on the global fit with a continuous evolution in the range $0<z<7$ also by \citet{kulkarni18} (models 1, 2, and 3).  All the surface densities are expressed in unit of $10^{-4} deg^{-2}$.\label{tab:surfdens}}
  \tablehead{
    \colhead{$i \le$} &
    \colhead{$\Sigma_{QSO}(2.5<z<3.0)$} &
    \colhead{$\Sigma_{QSO}(3.0<z<3.5)$} &
    \colhead{$\Sigma_{QSO}(3.5<z<4.0)$} &
    \colhead{$\Sigma_{QSO}(4.0<z<4.5)$} &
    \colhead{$\Sigma_{QSO}(4.5<z<5.0)$}
  }
  \startdata
  15.5 & 0.10-0.29 & 0.00-0.14 & 0.00-0.09 & 0.00-0.01 & 0.00-0.00 \\
  16.0 & 0.51-2.03 & 0.05-0.56 & 0.00-0.45 & 0.00-0.03 & 0.00-0.02 \\
  16.5 & 2.65-12.1 & 0.42-2.42 & 0.04-1.64 & 0.00-0.11 & 0.00-0.04 \\
  17.0 & 11.7-42.6 & 2.03-10.8 & 0.26-6.26 & 0.05-0.50 & 0.01-0.19 \\
  17.5 & 51.5-132.2 & 10.8-44.9 & 1.90-22.5 & 0.42-1.91 & 0.15-0.63 \\
  18.0 & 214.3-412.3 & 51.1-182.0 & 10.5-80.3 & 2.36-8.41 & 0.77-2.91 \\
  \enddata
\end{deluxetable*}
\begin{deluxetable*}{|c|c|c|c|c|c|}
  \tablecaption{Observed cumulative surface density of QSOs with $|b|> 25^\circ$ at different redshifts and $i$-band magnitudes (with $i \ge 15$).  In the Northern Hemisphere only the QSOs in the SDSS footprint have been considered, while in the South the known QSOs before the present survey in the Skymapper footprint from both \citet{DR14Q} and \citet{Veron10} have been considered.  All numbers are scaled to $10^4$ sq.deg. to allow a direct comparison with Tab.~\ref{tab:surfdens}.\label{tab:ObsQSOs}}
  \tablehead{
    \colhead{$i \le$} &
    \colhead{$\Sigma_{QSO}(2.5<z<3.0)$} &
    \colhead{$\Sigma_{QSO}(3.0<z<3.5)$} &
    \colhead{$\Sigma_{QSO}(3.5<z<4.0)$} &
    \colhead{$\Sigma_{QSO}(4.0<z<4.5)$} &
    \colhead{$\Sigma_{QSO}(4.5<z<5.0)$}
  }
  \startdata
  & North - South & North - South & North - South & North - South & North - South \\
  15.5    &    0.0 --   0.0   &    0.0 --   0.0   &    0.0 --   0.0   &    0.0 --   0.0   &    0.0 --   0.0 \\
  16.0    &    4.3 --   0.0   &    0.0 --   0.0   &    0.0 --   0.0   &    0.0 --   0.0   &    0.0 --   0.0 \\
  16.5    &    7.5 --   1.6   &    3.2 --   0.0   &    0.0 --   0.0   &    0.0 --   0.0   &    0.0 --   0.0 \\
  17.0    &   25.6 --  11.3   &    9.6 --   3.2   &    0.0 --   0.0   &    0.0 --   0.0   &    0.0 --   0.0 \\
  17.5    &   85.4 --  33.0   &   42.7 --  13.7   &    9.6 --   2.4   &    2.1 --   0.8   &    0.0 --   0.8 \\
  18.0    &  324.4 --  86.9   &  140.8 --  41.8   &   30.9 --   9.7   &    8.5 --   0.8   &    4.3 --   3.2 \\
\enddata
\end{deluxetable*}

\begin{figure*}
  \plotone{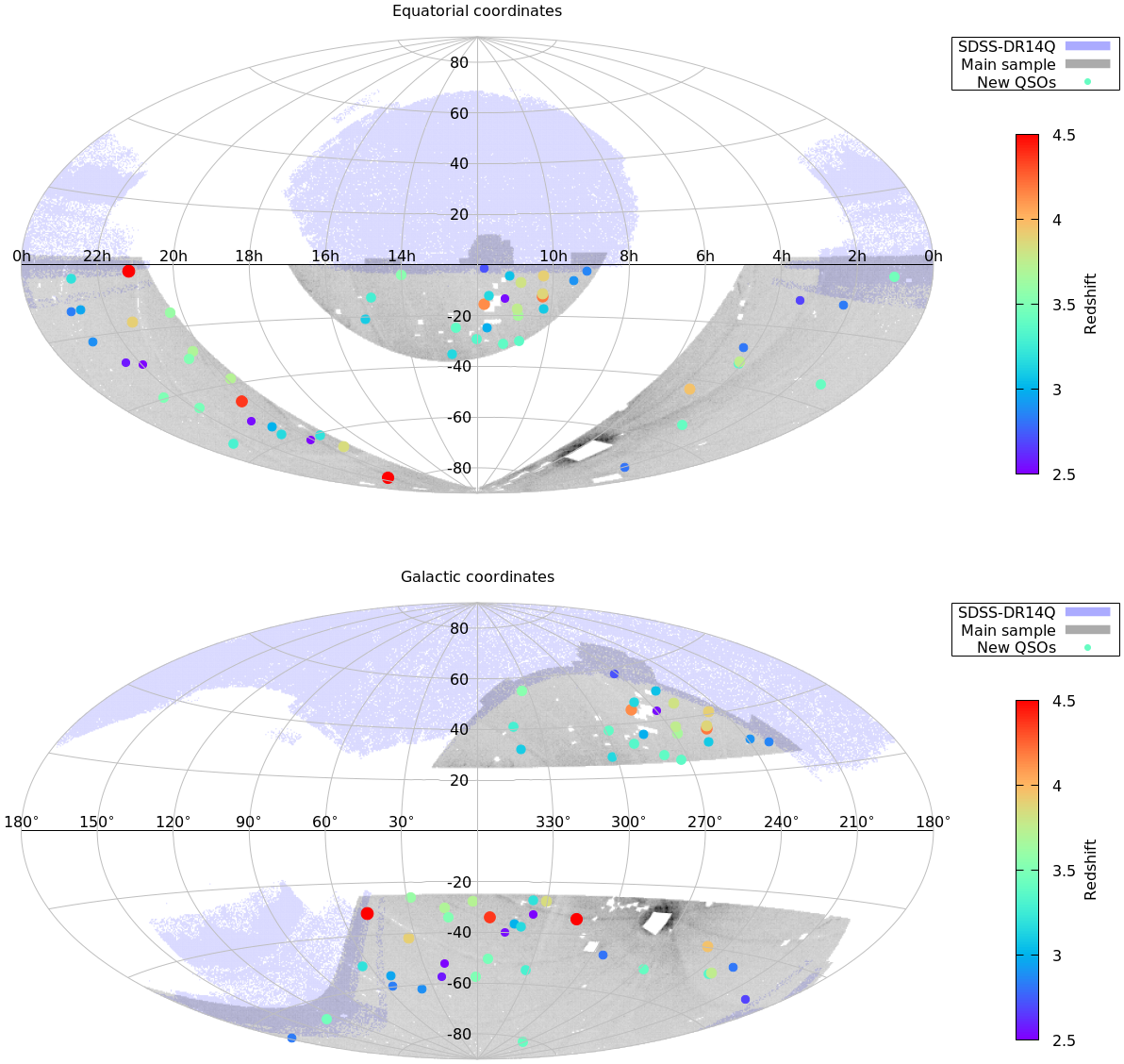}
  \caption{Maps of the locations of the sources in the {\it main sample} (\S\ref{sec:mainsample}, gray regions, a darker color indicates a higher density of sources).  The locations of the QSOs in the SDSS DR14Q \citep{DR14Q} are represented by blue shaded regions.  The new QSOs identified in this work are shown with filled circles, whose color indicates the redshift.  Upper panel: equatorial coordinates; lower panel: Galactic coordinates.}
  \label{Fig:skymap}
\end{figure*}
%

\section{The need and design of a new survey}

The typical range of apparent magnitudes of interest, having in mind a follow-up with high resolution spectroscopy, is $i \mincir 17$ at $z \sim 2.5$ and $i \mincir 18$ at $z \sim 4$.

In order to estimate the expected surface densities of QSOs, we have adopted the parameterization of the luminosity function by \citet{kulkarni18}. We extract random values of
redshift and $M_{1450}$ absolute magnitude following the best-fit luminosity functions in different redshift bins from Table 2 of \citet{kulkarni18}. Then we associate each simulated QSO to different templates from the Polletta empirical library of AGNs \citep{polletta} and convert the absolute magnitude into observed magnitudes in the adopted photometric system (i.e. Skymapper u,v,g,r,i,z; Gaia $BP$,G,$RP$; 2MASS J,H,K; WISE W1,W2,W3,W4; see next sections for a detailed description). We assume here a null Galactic dust extinction, since, as we discuss in the following, we select targets at high galactic latitudes.

Working at bright absolute magnitudes,  it is likely to be affected by small number statistics. To avoid this effect, we simulate a sky area of $10^5$ sq. deg., i.e.$\magcir 10$ times larger than the area available from the present survey, and repeat the simulation 10 times. This choice reduces the shot noise in the simulated number counts.

According to \citet{kulkarni18}, the best fit values of the QSO luminosity function (QLF) in their Table 2 can be affected by systematic errors due to the adopted survey selection functions. This is particularly true at z=2-4, where discontinuities and scatter in the QLF parameters appear over short redshift intervals. To avoid such discontinuities, we have computed the surface density adopting also the QLF resulting from a global continuous fit over the redshift range $0<z<7$ by \citet{kulkarni18} with a complex parameterization of the redshift evolution of the slopes, $\Phi^*$, and $M^*$ parameters. We used also their Models 1, 2, and 3 parameterizations to derive our estimates of the bright QSO number counts at $z>2.5$.

In Table \ref{tab:surfdens} we summarize the expected cumulative surface densities of QSOs in different bins of i-band magnitude and redshift. We provide the minimum and maximum expected values for the integral number counts based on the best fit values and on the models 1, 2, and 3 by \citet{kulkarni18}.
We do not consider here the effect of strong lensing, which can increase the luminosity of high-$z$ QSOs if their lines of sight are well aligned with the deep potential wells produced by galaxy over-densities or by single massive galaxies. The adopted luminosity functions, indeed, could be already affected by strong lensing in the bright end, especially at high-$z$ \citep{fan19,pacucci19}.

In the following, we will use our predictions in Table \ref{tab:surfdens} as a reference for the expected bright QSO number counts.  We expect that they should not be strongly affected by incompleteness, at least at very bright absolute magnitudes.

In Table \ref{tab:ObsQSOs} we compare the expected number of QSOs at galactic latitudes $| b | > 25^\circ $ with the number of presently known QSOs in the Northern and Southern hemisphere, respectively.  As already known, a significant discrepancy is present between the surface densities in the Northern and Southern hemispheres, in particular at $2.5\le z\le 4$. This is mainly due to the strong efforts, mainly by the Sloan Digital Sky Survey \citep{sdss}, devoted to the search for bright QSOs in the north. It is thus clear that a survey of bright high-$z$ quasars is still missing in the Southern Hemisphere.

Comparing the observed surface densities of bright QSOs at $z\ge 2.5$ in the North by Table \ref{tab:ObsQSOs} with the predicted ones in Table \ref{tab:surfdens}, it is clear that some of the models by \citet{kulkarni18} are underestimating the true number counts. In particular, their models 2 and 3 are predicting the lower boundaries in Table \ref{tab:surfdens}.  At completion, our survey will probably allow us to provide an assessment of the bright side of the quasar luminosity function at $z\ge 2.5$.

\section{A new selection of bright QSO candidates}

\subsection{The {\it main sample}}
\label{sec:mainsample}

\begin{figure}
\epsscale{1.2}
\plotone{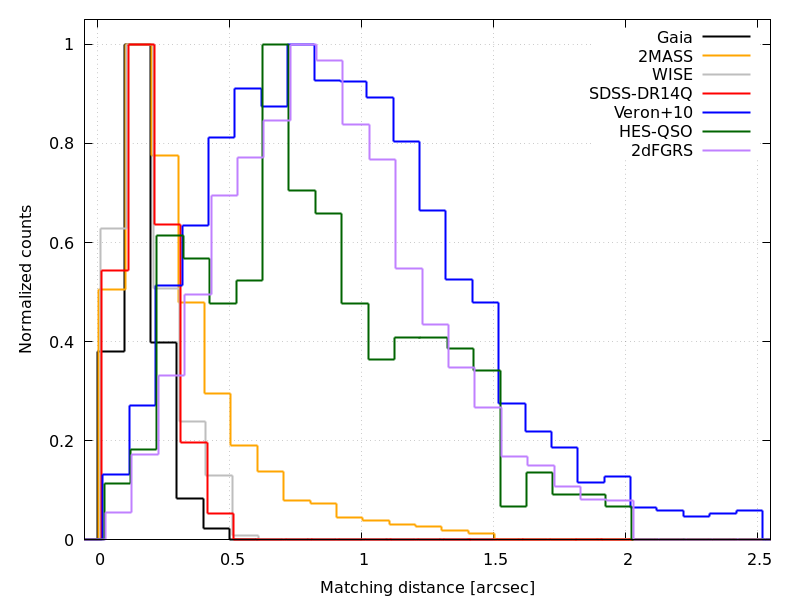}
\label{Fig:match_radius}
\caption{Histograms of the angular distances between the Skymapper sources in the {\it main sample} and the matched sources in the considered catalogs.  Each histogram has been normalized by its maximum in order to show all of them in the same plot.}
\end{figure}

In order to select new bright QSO candidates at redshift $\gtrsim 2.5$ in the Southern hemisphere (declination $<$~0$^\circ$) we have taken advantage of the following databases:
\begin{itemize}
    \item The Skymapper survey (DR1.1, \citealt{RefSkymapper});
    \item The Gaia DR2 data release (DR2, \citealt{RefGaiaMain}, \citealt{RefGaiaDR2});
    \item The WISE survey \citep{RefWise};
\end{itemize}
We considered all sources in the Skymapper survey with the following constraints:
\begin{enumerate}
    \item Galactic latitude $|b| > $~25$^\circ$;
    \item Magnitude in the $i$ band fainter than 14 and brighter than 18;
    \item Flags in the $i$ and $z$ bands equal to zero (i.e. availability of reliable $i$ and $z$ magnitudes);
    \item Availability of the Gaia magnitude in the $G$ band;
    \item Distance to the closest WISE source $<$~0.5'';
    \item Distance to the closest Gaia (DR2) source $<$~0.5'';
    \item Signal--to--noise ratio of the matching WISE  source in each of the first three bands $>$~3 (i.e. availability of reliable magnitudes in these bands).
\end{enumerate}
We limit our analysis to the magnitude range $i=$~14--18 in order to keep our samples as small as possible.  Besides, sources brighter than $i=14$ would hardly be high-$z$ QSOs, and sources fainter than $i>18$ are not interesting for our purposes. The constraint 3), in particular the request of a reliable $i$ magnitude, limits the effectiveness of this selection to redshifts $z \lesssim 5.3$. For the selection of higher redshift QSOs this constraint has to be relaxed.  We discarded the regions of the Large and Small Magellanic Clouds to avoid crowding and extinction. The initial sample (hereafter {\it main sample}) contains to \nmain{} objects spanning approximately 12,400 square degrees.  When available, we also collected the data in the following bands:
\begin{itemize}
    \item $u$, $v$, $g$, and $r$ from Skymapper;
    \item $BP$ and $RP$ magnitudes from Gaia;
    \item J, H, and K from the 2MASS \citep[][requiring a matching distance $<$~1.5'']{Ref2MASS}.
\end{itemize}
The angular distance matching radii for the above mentioned catalogs (and other reference catalogs, see below) have been determined by empirically checking the distributions of the angular separation histograms (see Fig.~\ref{Fig:match_radius}) with the aim of achieving the great majority of the true matches while minimizing the number of spurious associations.

\subsection{Source classification in the {\it main sample}}
\label{sec:classification}
In order to identify stars in the main sample we used the following criteria:
\begin{itemize}
    \item parallax (as measured by Gaia) significantly different from zero ($>3 \sigma$);
    \item Gaia proper motion along RA or DEC significantly different from zero ($>3 \sigma$).
\end{itemize}
\nstars{} objects out of \nmain{} (83.2\%) meet at least one of the two above criteria and in the following they will be considered as {\it bona fide} stars.

In order to identify known QSOs and extragalactic objects in the {\it main sample} we matched it against the following catalogs:
\begin{itemize}
    \item The SDSS DR14Q quasar catalog \citep[526,356 sources,][]{DR14Q} finding \nsdss{} matching entries within 0.5'';
    \item The 13th edition of the Veron--Cetty catalog \citep[167,566 sources,][]{Veron10}, finding \nveron{} matching entries within 2.5'' (only sources with a reliable spectroscopic redshift estimate have been considered);
    \item The 2dFGRS catalog \citep{2dfgrs}, finding \ntwodfgrs{} entries within 2'' (only sources with absorption spectra have been considered).
\end{itemize}
In this way we identified \nconfirmed{} spectroscopically confirmed QSO/AGN in the redshift range $0.005 < z < 5.06$, and \ngal{} sources with absorption spectra and no significant proper motion or parallax measurement, i.e., non-active galaxies (mainly at $z \lesssim 0.5$). In total, \nknown{} sources ($84.0\%$) in the {\it main sample} have a reliable object-type identification.  The remaining \nunk{} sources build up the {\it unknown} sample. 

\subsection{The {\it QSO candidate} sample}
\label{sec:cca}
In order to select new QSO candidates we need a method to identify the QSO characterizing properties among the \nunk{} sources in the {\it unknown sample}.  However, we have no access to their spectra, hence we must search for those properties in the available magnitudes.  Historically, this task has been accomplished by means of color selections (e.g. \citealt{2002-Richards-QSOSelection-SDSS}, \citealt{2013-Assef-SelectionWISE}, \citealt{2017-Tie-SelectionDES}), i.e. by cuts based on empirically identified linear combination of magnitudes (the so called {\it colors}.

\begin{figure*}[hbt!]
\epsscale{1}
\plotone{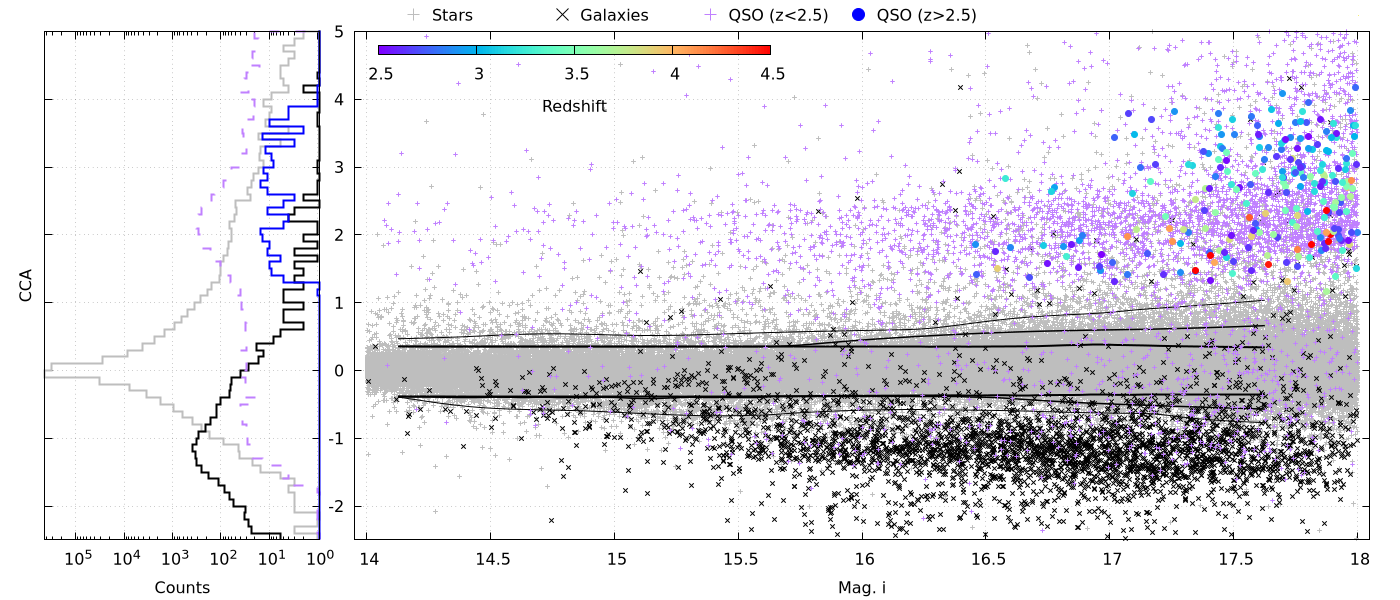}\\
\plotone{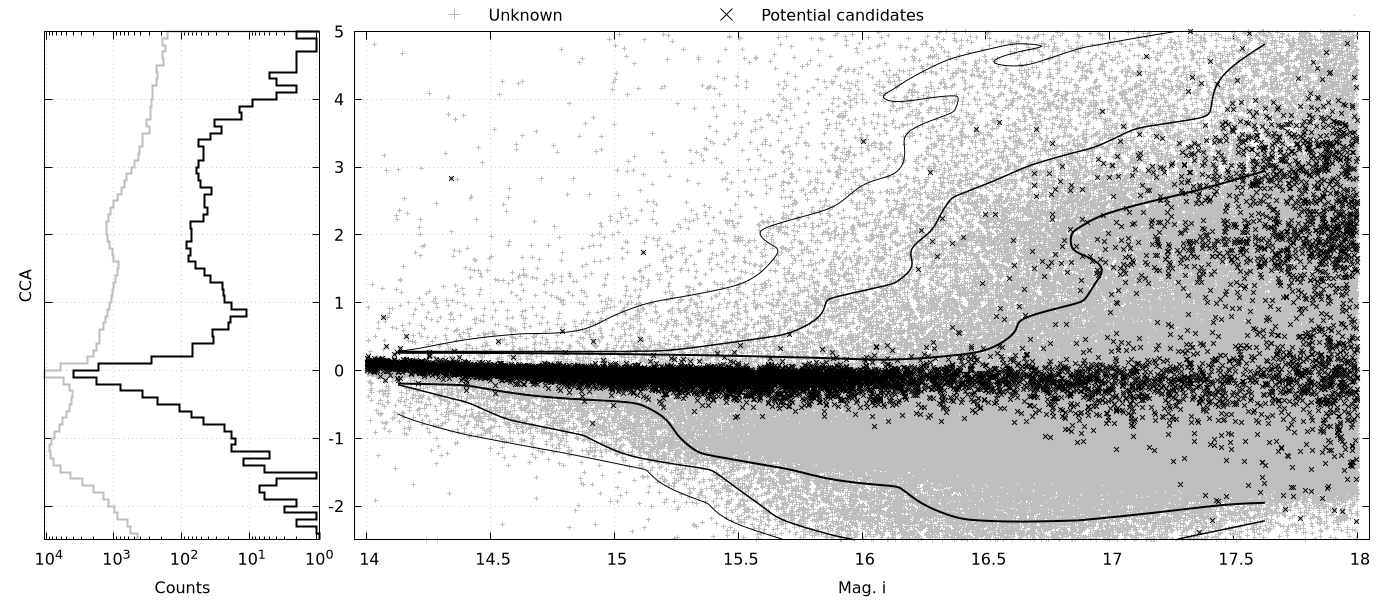}\\
\plotone{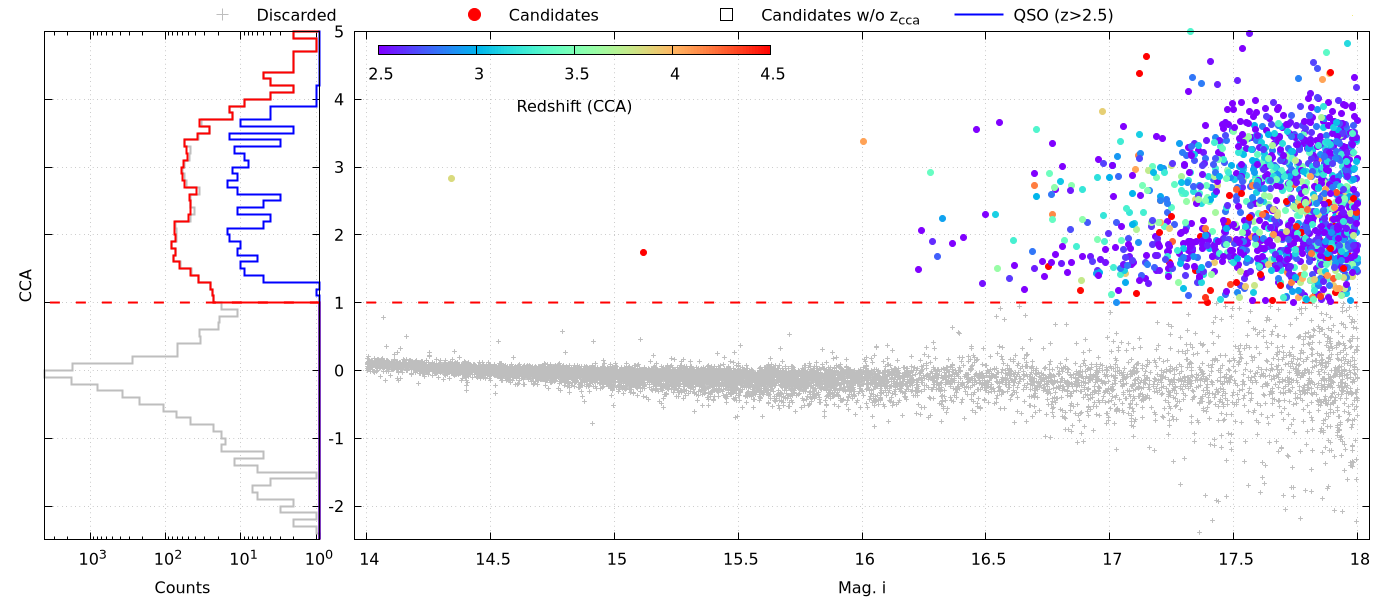}\\
\caption{The \CCA{}--$i$ mag. plane for the subsamples considered in this work. Upper panel: sources in the {\it main} sample for which a reliable type identification is available (\S\ref{sec:cca}). Stars are identified by gray ``+'' symbols, inactive galaxies by black cross symbols, low-$z$ ($<2.5$) QSOs with purple ``+'' symbols, high-$z$ ($>2.5$) QSOs with filled circles. The redshift for the confirmed QSOs with $z_{\rm spec} >2.5$ are shown with the color code shown in the colorbox in the upper left corner.  The inset on the left shows the histogram of the \CCA{} coordinate for the stars (gray), galaxies (black), low-$z$ QSOs (purple) and high-$z$ QSOs (blue). Middle panel: sources in the {\it main sample} without an object type identification (gray symbols, \S\ref{sec:cca}). The same sources after excluding extended (\S\ref{sec:extended}) and low-$z$ objects (\S\ref{sec:z_cca}) are highlighted in black, and represents potential high-$z$ QSO candidates. Lower panel: the {\it final} sample of high-$z$ QSO candidates, with the redshift $z_{\rm cca}$ estimated using the procedure described in \S\ref{sec:z_cca}.}
\label{Fig:cca1}
\end{figure*}

Here we follow a similar approach, but we identify the cuts in an automatic fashion using a machine learning procedure based on the Canonical Correlation Analysis\footnote{\url{https://en.wikipedia.org/wiki/Canonical_correlation}} \citep[CCA,][]{CCA}, rather then using color-color plots to isolate the interesting sources.  Our aim is to train the CCA using the object type classification (\S\ref{sec:classification}) as one of the canonical variables.  To this purpose we consider all the sources in the {\it main sample} with a clear object-type identification and attached a numerical label as follows:
\begin{itemize}
\item Label = -1: for the non-active galaxies;
\item Label = 0: for the stars;
\item Label = 2: for the spectroscopically confirmed QSOs with $z<2.5$;
\item Label = 3: for the spectroscopically confirmed QSOs with $z>2.5$.
\end{itemize}
This subset represents our {\it training sample}, and we use the numerical label as the first canonical variable\footnote{Canonical variables are obtained from input variables by means of a linear transformation. Since the numerical label is 1-dimensional it is by definition proportional to a canonical variable.}. The actual value of the numerical labels are rather arbitrary (up to constant scale factors and offsets). We just found a better separation when using a label for the stars sitting in the middle between inactive galaxies and QSO sources. Then we arranged the magnitudes discussed in \S\ref{sec:classification} in a matrix with as many rows as the number of sources, and as many columns as the available magnitude estimates, and apply the CCA procedure between this matrix and the numerical label discussed above.  The output of the CCA procedure is a linear transformation matrix which can be multiplied by the magnitude matrix to obtain a new, 1-dimensional coordinate (hereafter named \CCA{}), representing the canonical variable associated to magnitude estimates.  The CCA procedure ensures that the \CCA{} coordinate has the highest possible correlation with the numerical label, compatible with the data available in the training set. The \CCA{} coordinate for the sources in the {\it main sample} is shown in Fig.~\ref{Fig:cca1} (upper panel) as a function of the $i$ magnitude. Stars align across a rather narrow horizontal stripe at \CCA{}~$\sim$~0, while the non--active galaxies occupy the lower part of the plot.  Confirmed low-$z$ ($z < 2.5$) QSOs are spread throughout the whole \CCA{}--$i$ plane, but QSOs with $z>2.5$ cluster in the upper right corner, hence we expect new (i.e. not yet identified) QSOs at $z>2.5$ to be located in the same region.  

Then, we used the same transformation matrix used above to estimate the \CCA{} coordinate for the sources in the {\it unknown} sample, obtaining a \CCA{} value representative of the source object types (as was the case for sources in the training set).  We started our analysis by considering the sources with a reliable magnitude in all the above mentioned bands ($u$, $v$, $g$, $r$, $i$ and $z$ from Skymapper, $G$, $BP$ and $RP$ from Gaia, W1, W2 and W3 from WISE, J, H and K from 2MASS), then we proceeded analyzing the sources with incomplete photometric sets (i.e. with some magnitude missing among the 15 listed above).  Among the various configurations of bands we treated first the cases with the highest number of sources (allowing us to determine a more robust correlation), then the others with progressively less sources.  In each iteration we performed a linear fit against the original numerical label in order to renormalize the \CCA{} coordinates and span always the same dynamical range for each configuration of photometric bands.  In this way we have been able to compute the \CCA{} coordinates consistently for all the sources in the {\it main sample}.

Fig.~\ref{Fig:cca1} (middle panel, gray symbols) shows the location of the \nunk{} sources ($16 \%$) of the {\it unknown sample} remaining after the removal of the objects with a known object-type identification (stars, galaxies, QSOs). The $z>2.5$ QSOs we are looking for are expected to lie in the region at $\CCA \gtrsim 1$, but they are still confused in an overwhelming cloud of extended, inactive galaxies and low-$z$ AGN (shown in the upper panel of Fig.~\ref{Fig:cca1} with black "x" and purple ``+'' symbols, respectively).  It is therefore necessary to further distill our $z>2.5$ candidates by selecting against extended objects and low-$z$ sources.

\subsubsection{Excluding extended objects}
\label{sec:extended}

\begin{figure}[hbt!]
\epsscale{1.1}
\plotone{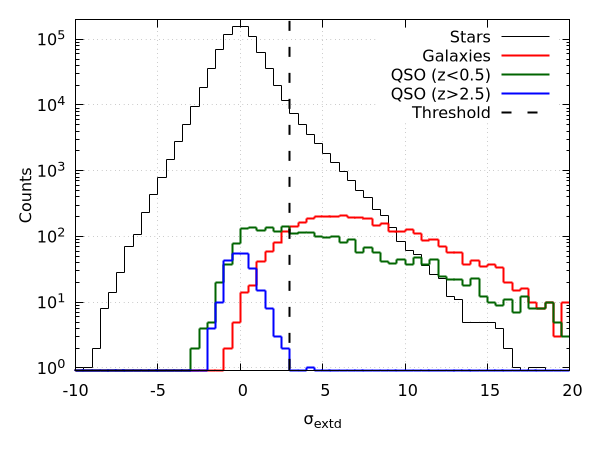}
\caption{Histogram of the $\sigma_{\rm extd}$ quantity (Eq.~\ref{eq:ext_z}) for all the sources in the {\it main} sample.  The threshold at $\sigma_{\rm extd}=3.0$ is shown with a vertical dashed line.  Sources above this threshold are assumed to be spatially extended and discarded.}
\label{Fig:etended_z}
\end{figure}

Since we are looking for bright high-$z$ QSOs we expect them to have a point-like appearance. In order to test whether an object in the {\it unknown} sample is spatially extended we have taken advantage of the comparison between the PSF and Petrosian magnitudes reported in the Skymapper catalog. The latter are supposed to be similar to the former only for point-like sources, while capturing more flux with respect to the PSF magnitudes for extended sources. In order to quantify such a difference we divided the whole {\it main sample} into bins of 0.1 mag and adopted the median of the differences between the PSF and the Petrosian magnitude as a reference value within each bin. Then, for each source, we interpolated the reference values corresponding to its PSF mags and computed the quantity:
\begin{equation}
\label{eq:ext_z}
\sigma_{\rm x,extd} = \frac{1}{2} ~ \sum_{x=i,z}
\frac{x_{\rm psf} - x_{\rm petro} - \langle x_{\rm psf} - x_{\rm petro} \rangle_{\rm ref}}{\sqrt{\sigma^2_{x_{\rm psf}} + \sigma^2_{x_{\rm petro}}}}
\end{equation}
where the $x$ represents either considered band, and $\sigma_{x}$ is the associated uncertainty\footnote{Note that in this equation $z$ refers to the magnitude in the $z$ band, not to the redshift.}.  We repeated the procedure in both the $i$ and in the $z$ bands and considered the average $\sigma_{\rm extd} = (\sigma_{\rm i,extd} + \sigma_{\rm z,extd}) / 2$ as an estimate of the significance of the object being extended.  The histogram of the values of $\sigma_{\rm extd}$ for the objects in the {\it main sample} with an available object-type identification is shown in Fig.~\ref{Fig:etended_z}.  Almost all confirmed QSOs with $z>2.5$ have $\sigma_{\rm extd} < 3$, hence we considered this value as a threshold to distinguish point--like sources from extended sources.  In this way we discarded 135,238 {\it bona fide} extended sources from the \nunk{} objects of the {\it unknown} sample ($81 \%$).

\subsubsection{Excluding probable low-$z$ ($z<2.5$) sources}
\label{sec:z_cca}
\begin{figure}[hbt!]
  \includegraphics[width=0.41\textwidth]{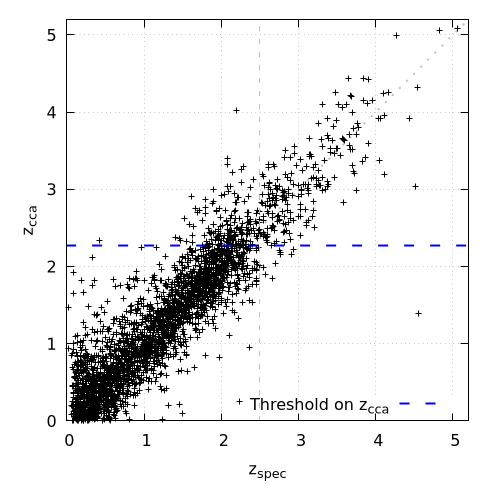}
  \includegraphics[width=0.41\textwidth]{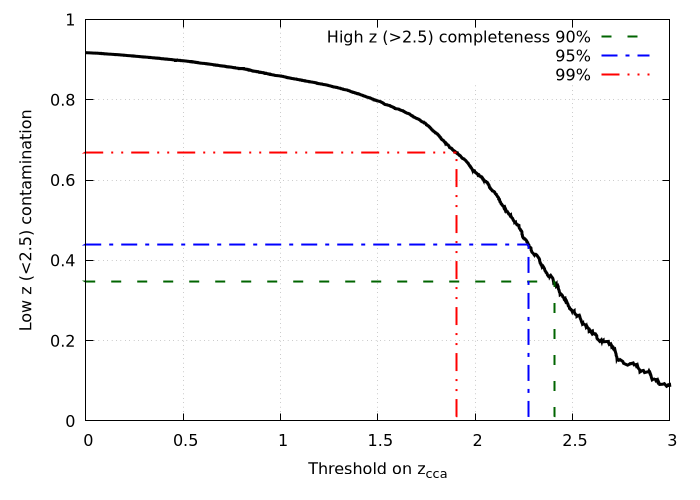}
  \caption{Upper panel: the $z_{\rm cca}$--$z_{\rm spec}$ correlation (scatter: $\sim$~0.36).  Lower panel: low-$z$ ($<2.5$) {\it ``contamination''} as a function of the adopted $z_{\rm cca}$ threshold and of the high-$z$ ($>2.5$) {\it ``completeness''} (dashed, dot--dashed and dot--dot--dashed lines).  We chose a $z_{\rm cca}$ threshold of 2.27 corresponding to a high-$z$ completeness of 95\% and an expected low-$z$ contamination in our {\it final} sample of $\sim$~44\%.}
\label{Fig:zcca}
\end{figure}

To estimate the redshift of the sources in the {\it main} sample we used again a CCA transformation, this time using the spectroscopic redshifts of the subsample of confirmed (\S\ref{eq:ext_z}) QSOs as a training set, and following the same procedure we used to calculate the \CCA{} coordinate.  The comparison between $z_{\rm cca}$ and $z_{\rm spec}$ for the confirmed QSOs with $\sigma_{extd} < 3$ is shown in Fig.~\ref{Fig:zcca} (upper panel). The scatter in the $z_{\rm cca}$ estimates is $\sim$~0.36.
Then we estimated the CCA redshift (hereafter $z_{\rm cca}$) using the resulting transformation matrix for the whole {\it unknown} sample.  To distinguish a low-$z$ ($z<2.5$) from an high-$z$ ($z>2.5$) source we calculated the following quantities:
\begin{itemize}
    \item {\it Low-$z$ ``contamination''}: ratio of the number of low-$z$ sources over the number of sources with $z_{\rm cca}$ above a given threshold;
    \item {\it High-$z$ ``completeness''}: ratio of high-$z$ sources with $z_{\rm cca}$ above a given threshold, over the total number of high-$z$ confirmed QSOs.
\end{itemize}
A plot of these quantities, for all the possible value of the $z_{\rm cca}$ threshold, is shown in Fig.~\ref{Fig:zcca} (lower panel, black solid line): a high-$z$ {\it ``completeness''} of 95\% can be reached with a threshold at $z_{\rm cca} = 2.27$, corresponding to a low-$z$ {\it ``contamination''} of 44\% (dot-dashed blue line). Increasing the high-$z$ {\it ``completeness''} to 99\% (blue dot-dot-dashed) would yield a much higher contamination, while decreasing to 90\% (green dashed line) would yield only a small improvement in contamination.  Hence, we chose $z_{\rm cca} = 2.27$ as discriminating threshold to select against low-$z$ QSO candidates.

\subsubsection{The final QSO candidate sample}
\label{sec:candidate_sample}
By discarding the extended and low-$z$ sources from the \nunk{} objects of the {\it unknown} sample, we are left with 11,598 potential QSO candidates (black cross symbols in the middle panel of Fig.~\ref{Fig:cca1}).  Besides excluding a significant fraction of sources in the {\it unknown} sample (93\%), the above procedure allowed to obtain a better separation of the remaining sources in the \CCA{}--$i$ mag plane, resulting in an increased contrast between the peaks above and below \CCA{} $\sim$~1 in the histogram on the left of middle panel in Fig.~\ref{Fig:cca1}, and suggesting that a threshold on the \CCA{} value might allow to exclude the non--QSO sources.   As discussed above, the group at \CCA{}~$>$~1 is likely associated with high-$z$ QSOs, while the group at \CCA{} $<$~1 is associated with stars and inactive galaxies.
Therefore we discarded all the source with \CCA{} $<$~1 to obtain a {\it final} sample of \nfinal{} high-$z$ QSO candidates.  The lower panel of Fig.~\ref{Fig:cca1} shows the location of such candidates in the \CCA{}--$i$ mag. plane, and their expected redshift (color-coded, as calculated in \S\ref{sec:z_cca}).

As a consistency check, we note that all the known QSOs with $z>2.5$ (blue line in the histogram, both in upper and lower panel of Fig.~\ref{Fig:cca1}) lie above the adopted \CCA{} threshold, as expected. 

The number of DR14Q and Veron sources with $z>2.5$ in our main sample is 68. Considering that the Skymapper footprint is ~8.3 times larger than the previously surveyed area, we extrapolate a number $\sim 564$ new QSOs with $z>2.5$ in our {\it QSO candidate} sample.  Given the size of the QSO candidate sample (\nfinal{} sources) we expect a lower limit for the success rate for high-$z$ ($>2.5$) QSO identification of $\sim$~40\%. Actually, the fraction of new high-$z$ ($>2.5$) QSO spectroscopically confirmed among the candidates we could observe (\S\ref{sec:observations}) is $\sim$~80\%.

At this stage we can also compute the fraction of DR14Q and Veron QSOs with $z>2.5$ and $i < 18$ satisfying all the conditions to be selected by our procedure, $93\%$, and use it as an indication of the completeness of our QSO sample.  The $7\%$ of the known QSOs lost were sources with a predicted CCA redshift below our threshold of $z_{\rm cca} = 2.27$ (\S\ref{sec:z_cca}).

\section{Spectroscopic confirmations}  
\label{sec:observations}

In order to validate the above-described selection criteria (and test variants), we have carried out extensive spectroscopy observations at Las Campanas Observatory and at the ESO-NTT telescope at La Silla.  The first pilot study has been carried out at the Magellan telescopes in 2018 using LDSS-3 (Clay Telescope) and IMACS (Baade Telescope).  Observations were obtained in various nights during bright time and variable weather conditions.  With LDSS-3 the VPH-all grism has been used with the 1"-central slit and no blocking filter, covering a wavelength range between 4000 - 10000 {\AA} with a low resolution of R $\sim$800.
With IMACS, we used the \#300 grism with a blaze angle of 17.5deg, covering a wavelength range between 4000 - 10000 {\AA} with a dispersion of 1.34 \AA/pixel. Based on these first results, the selection technique has been adjusted in order to include candidates at higher redshift. 
In February 2019 we were awarded 2 nights at the du Pont telescope to validate the optimized criteria, and we observed several new candidates with the Wide Field CCD (WFCCD) blue grism that covers a wavelength range between 3700 - 8000 {\AA} providing a 2 \AA /pixel dispersion.  

The NTT spectroscopic campaign has been carried out during the ESO observing period P103 under the proposal 0103.A-0746 (PI. A. Grazian).  Three nights of spectroscopy have been executed during 27-30 April 2019.  The EFOSC2 instrument was used, equipped with the grism \# 13 (wavelength range $\lambda\sim 3700-9300$ {\AA}).  Since our main targets are relatively sparse in the sky, we carried out long-slit spectroscopy with exposure times between 3 and 7 minutes per object.

Finally, in June 2019 we performed a few exposures at TNG (La Palma) using the Low Resolution Spectrograph (Dolores) with the LR-B grism (resolution $\sim$~600), a 1" slit aperture and an exposure time of 10 minutes per object, in order to validate our selection criteria against low-$z$ AGNs (\S\ref{sec:z_cca}).

\begin{figure}
\epsscale{1.25}
\plotone{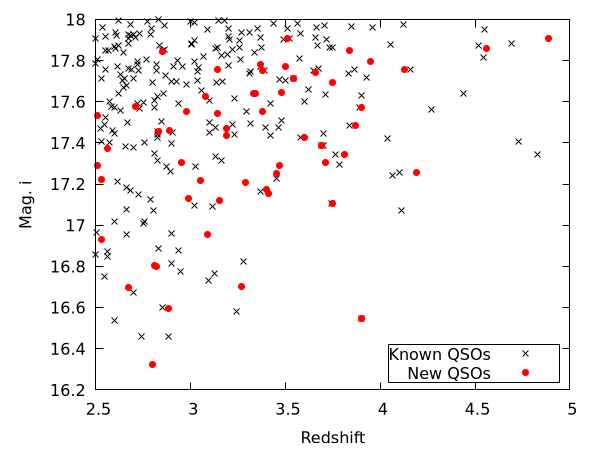}
\caption{The redshift-$i$ magnitude plane of the QSOs in the area of the present survey. Black crosses: QSOs known before the present observations; Red filled circles: new spectroscopic redshifts obtained in the present survey.}
\label{Fig:i_vs_z}
\end{figure}

In total we observed \nOBSfin{} sources from our {\it final}  QSO candidate sample (\S\ref{sec:candidate_sample}) of \nfinal{} sources.  Among these sources, \nFINconf{} turned out to be genuine high-$z$ QSOs with $z>2.5$, 12 are low-$z$ QSOs with $z<2.5$ and 1 is a star.  The details of the candidate observations are summarized in Tab.~\ref{tab:ObservCandidates}, while Fig.~\ref{Fig:i_vs_z} shows the redshift-$i$ magnitude plane of the newly discovered QSOs (red circles) and the known QSOs before this work (black cross symbols).  So far, we achieved a success rate in identifying new high-$z$ QSOs sources of $\sim$~80\%.  On the other hand, we observed only a small fraction of the total QSO candidate sample (\nOBSfin/\nfinal{} sources, $\sim$~5\%), hence the success rate may be biased by the choice of the most promising candidates for the observations.

In the early phases of the project we experimented different versions of the selection algorithm and tested its limits and characteristics with the pilot spectroscopic runs and in part of the NTT run.
As a consequence, we also observed sources that do not belong to the {\it final} sample.
For completeness, we report the observation details for these \nADD{} additional sources in Tab.~\ref{tab:ObsNonCandidate}.  Among them we found: 2 QSOs at $z>2.5$, 50 low-$z$ QSOs, and 15 non-QSO sources.  The two QSOs at  $z>2.5$ were not selected in the {\it main} sample because one has an $i$ magnitude fainter than the threshold of $i=18$ and the other has a Skymapper astrometric position differing of more than $0.5"$ from Gaia DRS2, probably due to image defects in the Skymapper data, as we checked with Skymapper cutouts.

Further observing runs at the DuPont and NTT telescopes have been approved in order to expand our spectroscopically observed sample.  All the details and results of the  spectroscopic runs will be described in a future paper.

\section{Conclusions}

The aim of the present project was to identify new, bright ($i<$~18) QSOs at relatively high redshift ($z>2.5$) in the Southern Hemisphere with a high success rate.  At this stage completeness represented a secondary requirement.

Finding in an efficient way relatively high-redshift QSOs is a kind of a {\it needle in a haystack} task.  Our approach has been to take advantage of large high-quality photometric and astrometric databases provided by Skymapper, WISE, 2MASS and Gaia, in order to remove sources identified with high-reliability as contaminants (stars, low-$z$ QSOs and galaxies).  Then, with the help of a Canonical Correlation Analysis \citep{CCA} we have selected among the remaining {\it unkown} objects a sample of \nfinal{} $z>2.5$ QSO candidates, whose completeness is also expected to be high ($\magcir 90 \%$ for objects up to $z \sim 5$), estimated on the basis of the number of known QSOs that the method would select.

Indeed, the first campaigns of spectroscopic confirmations have been characterized by a high success rate ($\sim$~81\%), and already at this preliminary stage the number of bright QSOs in the southern emisphere has been significantly increased, as shown in Fig.~\ref{Fig:i_vs_z}.  The new \nFINconf{} (Tab.~\ref{tab:ObservCandidates}) plus 2 (Tab.~\ref{tab:ObsNonCandidate}) QSOs with $z>2.5$ and $i<18$ are now available to the astronomical community  for high-resolution spectroscopic follow-up and the studies of cosmology and fundamental physics described in the introduction.

We are continuing our campaigns of spectroscopic confirmations and at the same time we are exploring other statistical techniques in addition to the CCA analysis to further improve the properties of the selection and extend its range of applicability.

\acknowledgments
We thank Luca Pasquini and Carlos Martins for enlightening discussions.
This work is based on data products from observations made with ESO Telescopes at La Silla Paranal Observatory under ESO programme ID 103.A-0746(A).
The national facility capability for SkyMapper has been funded through ARC LIEF grant LE130100104 from the Australian Research Council, awarded to the University of Sydney, the Australian National University, Swinburne University of Technology, the University of Queensland, the University of Western Australia, the University of Melbourne, Curtin University of Technology, Monash University and the Australian Astronomical Observatory. SkyMapper is owned and operated by The Australian National University's Research School of Astronomy and Astrophysics. The survey data were processed and provided by the SkyMapper Team at ANU. The SkyMapper node of the All-Sky Virtual Observatory (ASVO) is hosted at the National Computational Infrastructure (NCI). Development and support the SkyMapper node of the ASVO has been funded in part by Astronomy Australia Limited (AAL) and the Australian Government through the Commonwealth's Education Investment Fund (EIF) and National Collaborative Research Infrastructure Strategy (NCRIS), particularly the National eResearch Collaboration Tools and Resources (NeCTAR) and the Australian National Data Service Projects (ANDS)
This work has made use of data from the European Space Agency (ESA) mission {\it Gaia} (\url{https://www.cosmos.esa.int/gaia}), processed by the {\it Gaia} Data Processing and Analysis Consortium (DPAC, \url{https://www.cosmos.esa.int/web/gaia/dpac/consortium}). Funding for the DPAC has been provided by national institutions, in particular the institutions participating in the {\it Gaia} Multilateral Agreement.
This publication makes use of data products from the Two Micron All Sky Survey, which is a joint project of the University of Massachusetts and the Infrared Processing and Analysis Center/California Institute of Technology, funded by the National Aeronautics and Space Administration and the National Science Foundation
This publication makes use of data products from the Wide-field Infrared Survey Explorer, which is a joint project of the University of California, Los Angeles, and the Jet Propulsion Laboratory/California Institute of Technology, funded by the National Aeronautics and Space Administration.
This paper includes data gathered with the 6.5 meter Magellan Telescopes located at Las Campanas Observatory, Chile.
We thank Societ\`a Astronomica Italiana (SAIt), Ennio Poretti, Gloria Andreuzzi, Marco Pedani, Vittoria Altomonte and Andrea Cama for the observation support at TNG.  Part of the observations discussed in this work are based on observations made with the Italian Telescopio Nazionale Galileo (TNG) operated on the island of La Palma by the Fundación Galileo Galilei of the INAF (Istituto Nazionale di Astrofisica) at the Spanish Observatorio del Roque de los Muchachos of the Instituto de Astrofisica de Canarias.

\vspace{5mm}
\facilities{Skymapper, Wise, 2MASS, Gaia, Magellan:Baade (IMACS), Magellan:Clay (LDSS-3), du Pont (WFCCD), TNG (Dolores)}


\startlongtable
\begin{deluxetable*}{r|c|c|c|l|c|c|c|c}
  \tablecaption{List of Skymapper sources in the {\it Candidate} sample observed in our campaigns.\label{tab:ObservCandidates}}
  \tablehead{
 \colhead{Skymapper ID}   & 
 \colhead{R.A. (J2000)}   & 
 \colhead{Decl. (J2000)}  & 
 \colhead{Date Obs.}      &
 \colhead{$m_i$}          &
 \colhead{Obj. type}      &
 \colhead{$z_{\rm spec}$} &
 \colhead{Instrument}     & 
 \colhead{Notes} 
  }
  \startdata
  7683342        &   00:35:58.10  &  -20:05:56.25  &  2018-11-22  &  16.842   &  QSO     &  1.53     &  LDSS-3      &                   \\
  ...            &   ...          &  ...           &  ...         &  ...      &  ...     &  ...      &  ...         &                   \\
  \enddata
\end{deluxetable*}

\startlongtable
\begin{deluxetable*}{r|c|c|c|l|c|c|c|c}
  \tablecaption{List of Skymapper sources observed in our campaigns which are not in the final {\it Candidate} sample.\label{tab:ObsNonCandidate}}
  \tablehead{
 \colhead{Skymapper ID}   & 
 \colhead{R.A. (J2000)}   & 
 \colhead{Decl. (J2000)}  & 
 \colhead{Date Obs.}      &
 \colhead{$m_i$}          &
 \colhead{Obj. type}      &
 \colhead{$z_{\rm spec}$} &
 \colhead{Instrument}     & 
 \colhead{Notes} 
  }
  \startdata
  9163238    &   02:00:16.25  &  -06:52:09.06  &  2018-11-23  &  16.688   &  QSO     &  1.75     &  LDSS-3       &             \\
  ...        &   ...          &  ...           &  ...         &  ...      &  ...     &  ...      &  ...          &              \\
\enddata
\end{deluxetable*}

\end{document}